\documentclass{aa}
\usepackage{graphicx}

\def\lsun{L_\odot}
\def\msun{M_\odot}
\def\be{\begin{equation}}
\def\ee{\end{equation}}
\def\gtsima{$\; \buildrel > \over \sim \;$}
\def\simgt{\lower.5ex\hbox{\gtsima}}
\def\ltsima{$\; \buildrel < \over \sim \;$}
\def\simlt{\lower.5ex\hbox{\ltsima}}

\begin{document}
   \title{Synchrotron emission from circumstellar disks around
          massive stars}


   \author{Yu. A. Shchekinov
          \inst{1},
          A. M. Sobolev
          \inst{2}
          }

   \offprints{Yu. Shchekinov}

   \institute{Department of Physics, Rostov State University,
              Rostov on Don 344090, Russia,\\
              and Isaac Newton Institute of Chile, Rostov on Don Branch\\
              \email{yus@phys.rsu.ru}
         \and
              Astronomical Observatory, Ural State University,
              Ekaterinburg 620083, Russia\\
              \email{Andrej.Sobolev@usu.ru}
             }

   \date{Received Nov 13, 2002; accepted Dec 13, 2002}

   \abstract{We argue that the interaction of stellar wind with the surface
of a circumstellar accretion (or protoplanetary) disk can result in
the acceleration of relativistic electrons
in an external layer of the disk, and produce synchrotron radiation.
Conservative estimates give a total synchrotron luminosity $L_s\sim
10^{-5}\lsun$ for a central star with $\dot M=10^{-6}\msun$ yr$^{-1}$,
comparable with the value observed around the TW object in the W3(OH) region.
   \keywords{accretion disks -- stars: circumstellar matter --
   interstellar masers
               }
   }
\authorrunning{Shchekinov \& Sobolev}
\titlerunning{Synchrotron from circumstellar disks}
   \maketitle
%

\section{Introduction}

Recent investigations have shown that H$_2$O masers in the vicinity
of young stellar objects can be associated both with jets and
circumstellar disks (see Torrelles et al. \cite{torrelles} for the
most recent review). It is suggested that systems in which H$_2$O
masers trace disks are less evolved than those in which masers
trace outflows.

The detection of synchrotron emission at $\nu=1.6-14.7$ GHz from a very
young massive stellar object associated with H$_2$O masers in the
W3(OH) region (Turner \& Welch 1984, hereafter TW object)
demonstrated the presence of relativistic electrons in the region
nearly coincident with the location of the H$_2$O masers (Reid et al.
\cite{reid}). These authors interpreted the synchrotron emission as
related to acceleration of the relativistic electrons in shock
waves associated with a powerful jet with E-W orientation
emanating from the young stellar object (Reid et al. \cite{reid},
Wilner, Reid, \& Menten \cite{wilner}). Association with the jet
is based on the measurements of proper motions of the H$_2$O masers
which are compatible with the hypothesis of a bipolar outflow moving
in the E-W direction (Alcolea et al. \cite{alcolea}).
We argue that interpretation of proper motions pattern of the H$_2$O maser
by Alcolea et al. (\cite{alcolea}) is not
unique (a similar situation is described in Fiebig et al.
\cite{fiebig96}) and does not rule out the hypothesis of formation
of the H$_2$O masers on the surface of the circumstellar
disk or in a molecular ring seen edge-on. Evidence pro and contra
the disk hypothesis are presented throughout paper.

Interstellar H$_2$O masers are usually located in regions
influenced by strong MHD shocks emanating from young
stellar objects. These shocks affect the boundaries of the jets,
outflows, the surface of circumstellar disks and the parts of the
ambient molecular core that are closest to the young stellar
object. So one might, in principle, expect that the acceleration
of relativistic electrons generating synchrotron emission can
arise not only at the jet boundaries and can be a common
phenomenon for the vicinities of H$_2$O masers. The question therefore
arises whether synchrotron emission at locations close to H$_2$O
masers does necessarily originate in jets or if an alternative
explanation connected with high energy processes in circumstellar
disks is possible.

In this paper we propose estimates which show that the interaction
of stellar wind from a central star with an accretion disk can
support favorable conditions for the acceleration of electrons and
the generation of synchrotron radiation from relatively narrow
boundary layers on the disk surface. We show that this interaction
can produce a total synchrotron luminosity in the TW object comparable
to what is observed. In Section 2 we present simple estimates of
the acceleration mechanism, and calculate the resulting
synchrotron emission from the disk surface, in Section 3 we
describe the spatial distribution of the synchrotron radiation, in
Section 4 we provide arguments in favor of a circumstellar disk
around the TW object, Section 5 summarizes the results.


\section{Energy requirements}

Strong winds from early-type stars generate through hydrodynamical
instabilities numerous shocks with Mach numbers up to 20 (Carlberg
\cite{carl}, Owocki \& Rybicki \cite{owocki}). The density of shocks
increases near the disk surface due to the Kelvin-Helmholtz (KH) instability
of the shear flow between the wind and the surface, and
subsequent interaction of the wind with the vortices and coherent
structures generated by the instability. On nonlinear stages these
structures have sizes and spacing of the order of the thickness of
the mixing layer (Roshko \cite{roshko}, Head \& Bandyopadhyay
\cite{head}): $\delta\sim \sqrt{4\eta r/v_\infty}$.
Assume that the energy density of the shocks in a layer of thickness
$\Delta H(r)$ is a fraction $\beta$ of the wind energy density:
$\varepsilon_s=\beta\rho v_\infty^2/2$, where
\be
\label{density}
\rho={\dot M\over 4\pi r^2v_\infty},
\ee
$v_\infty$ is the terminal velocity of the wind, $\dot M$, the mass loss
rate. The total energy of shocks in the boundary layers from both sides of
the disk is

\be
E_s={\beta\dot Mv_\infty\over 2}\int\limits_{R_\ast}^R{\Delta H(r)\over r}dr,
\ee
where $R_\ast$ is the stellar radius, $R$, the disk radius. Assume that
$\Delta H(r)=hr$, $h\ll 1$. Then

\be
E_s= 10^{44}\beta h\dot M_{-6}v_{8}R_{3}\, {\rm erg},
\ee
where $\dot M_{-6}=\dot M/(10^{-6}\msun~{\rm yr}^{-1})$,
$v_{8}=v_\infty/(10^8~{\rm cm~s^{-1}})$, $R_{3}=R/(10^3~{\rm AU})$
(note that the radius of the synchrotron structure observed by Reid et. al.
\cite{reid} is of the order of $10^3$ AU).

Charged particles will be accelerated by the shocks
through the first order Fermi mechanism. Assume that the mechanical energy of
shock waves is transformed into cosmic ray energy until the nonlinear back
reaction of accelerated particles leads to significant dissipation of the
large-scale turbulence (Zank, Axford, \& McKenzie 1990). Under these
conditions one can expect that in the steady state the energy accumulated
in non-thermal ions will be comparable to the energy stored in shocks, while
the fraction of energy contained in relativistic
electrons is $(m_e/m_p)^{(q-1)/2}$, $q$ being the exponent of a power law 
spectrum of relativistic electrons (see below). 
The maximum energy of non-thermal ions
$e_m$ is found from the condition (Bykov \& Fleishman 1992)

\be
{3ul_s\over v\Lambda}\sim 1,
\ee
where $u$ is the shock speed, $l_s$ is the distance between the shocks,
$v$ is the particle velocity, $\Lambda$ is the transport length (Toptygin
1985, Bykov \& Fleishman 1992)

\be \Lambda\simeq 0.3l_s(r_b/l_s)^{2-\xi}, \ee where $r_b=pc/eB$,
$\xi$ is the power-law exponent of the fluctuating magnetic field
spectrum $d\langle b^2\rangle/dk\sim k^{-\xi}$ at $kl_s\gg 1$
(Bykov \& Fleishman 1992). The mean distance between the shocks in
strong radiation-driven winds is of the order $l_s\sim
10^{-2}R_\ast(r/R_\ast)^2$ (Lucy 1982, White 1985), and can reach
$l_s\sim 0.1-10$ AU at $r\sim 1-10$ AU for an O- or early
B-type star. The actual value can be smaller if multiple reflections
of the shocks on the vortices and coherent structures in the
boundary layer of the disk are taken into account. As a
conservative estimate one can take $l_s\sim 1$ AU. This results in
$e_m\sim 300$ GeV for $\xi=1.5$, and $e_m\sim 10$ GeV for
$\xi=1.7$. We will assume that the energy spectrum of non-thermal
electrons extends also to these values, so that the characteristic
frequency $\nu_b=2\pi \gamma eB/m_ec^2$, where $\gamma=e/m_ec^2$,
can be as high as 20 GHz.

With these assumptions and for
a power-law spectrum of relativistic electrons

   \begin{equation}
   \label{spect}
      N(e)de=\kappa e^{-q}de, \, e\geq e_{\mathrm min}, \, q>2,
   \end{equation}
where $\kappa=(q-2)\varepsilon_s (m_e/m_p)^{q-2/2}e_{\mathrm min}^{q-2}$,
with $q=2.2$ as observed in the TW object and $e_{\mathrm min}=\gamma_{\mathrm 
min} m_ec^2$, one can obtain the total synchrotron luminosity

\be \label{lumin} L_{sD}=\int\limits_{}^{14.7~{\rm GHz}}
L_s(\nu)d\nu= 10^{-3}\beta h\gamma_{min}^{0.2}\dot
M_{-6}v_{8}R_{3} B_{0.01}^{1.6} \, \lsun, \ee where
$B_{0.01}=B/(0.01~{\rm G})$. The synchrotron luminosity of the TW
object in the frequency range $\nu= 1.6-14.7$ GHz is $L\cong
5\times 10^{-6}\, \lsun$, as inferred from Reid et. al
(\cite{reid}). It is seen that at fiducial values of the
parameters $\beta h\sim 5\times 10^{-3}\gamma_{min}^{-0.2}$
provides a synchrotron luminosity from the boundary layer comparable
to the luminosity observed in the TW object. Note, that with the same
assumptions about the generation of relativistic electrons in the jet
with a collimation angle $h$ its total synchrotron luminosity is
$L_{sJ}=hL_{sD}\ll 1$, so that the observed luminosity of the TW
object can be understood only if $\beta h^2\sim 5\times
10^{-3}\gamma_{min}^{-0.2}$, i.e. under the jet hypothesis a much
larger (by a factor $1/h\gg 1$) fraction of the wind energy
should be transformed into the shocks which accelerate
electrons.

\section{Angular size versus frequency}

Reid et al. (\cite{reid}) found that the angular size of the synchrotron
source decreases with frequency, and attributed it to self-absorption of
synchrotron photons. In our model the maximal (along
radius) optical depth is

\be
\tau=2\int\limits_{r_0}^R \mu(r) dr,
\ee
where $r_0$ is the inner disk radius, $\mu$ is the absorption coefficient
for synchrotron emission (see Ginzburg 1975)

\be \label{absorption}
\mu=g(q){e^3\over 2\pi m_e}\left({3e\over 2\pi m_e^3c^5}\right)^{q/2}
\kappa B^{(q+2)/2}\nu^{-(q+4)/2},
\ee
$g(q)\simeq 0.7$ for $q\simeq 2$ (see Ginzburg \cite{ginzburg}). This
gives for the above assumption about the energy density in relativistic
electrons $\varepsilon(r)=\beta \dot M v_\infty/8\pi r^2$, and
for $B=B_0(r_0/r)^b$ optical depth

\be
\tau=3\times 10^8\beta B_{0,0.01}^{(q+2)/2} \dot M_{-6}r_0^{-1},~
{\rm at~ \nu=10~GHz},
\ee
and

\be \tau=4\times 10^{11}\beta B_{0,0.01}^{(q+2)/2} \dot M_{-6}
r_0^{-1},~{\rm at~ \nu=1~GHz}, \ee and obviously $\tau$ is much less than
one for $r_0>R_\ast$. One can show that free-free absorption is
also negligible for the parameters of interest. Note that for the
best fit parameters derived by Reid et al. within the jet model
($e_{min}=15m_ec^2$, the total density of relativistic electrons
$\int N(e)de=5\times 10^{-3}$ cm$^{-3}$, $B_0=0.01$ G) Eq.
(\ref{absorption}) gives optical depth $\tau\sim 10^{-22}L\ll 1$
at $\nu=1$ GHz, $L$ being the characteristic size of the absorbing
region. Under such conditions the decrease of the angular
size of the source with frequency (approximately as $\theta\propto
\nu^{-1}$, Reid et. al. 1995) can be explained only
by assuming that the spectral properties of the synchrotron
emission vary with radius. For a power-law spectrum of
relativistic electrons as given by Eq. (\ref{spect}) in the energy
interval $e=[e_{min},e_{max}]$ the synchrotron emissivity is
approximately a power-law $I_\nu\propto \nu^{-(q-1)/2}$ in the 
frequency interval $\nu=0.29[\nu_{min},\nu_{max}]$, where
$\nu_{min;max}$ correspond to minimal and maximal energies of
electrons. Below and above this interval the emissivity varies as
$I_\nu\propto (\nu/\nu_{min})^{(3q-1)/6}$, and $I_\nu\propto
(\nu/\nu_{max})^{(q-2)/2}\exp(-\nu/\nu_{max})$, respectively. The
mean frequency in the spectrum is determined by $q$, $\nu_{min}$
and $\nu_{max}$, and $\bar\nu=(3-q)\nu_{max}/(5-q)$ for $q<3$ and
$\nu_{max}\gg \nu_{min}$. The observed size of a synchrotron
source will then decrease with frequency if the mean frequency in
the spectrum $\bar\nu$ decreases with radius: at fixed frequency
$\nu$ the main contribution to the observed flux comes from a
region with radius corresponding to the condition $\nu <
0.29\nu_{max}(r)$.

For fixed boundary energies $e_{min}$ and $e_{max}$ in the
spectrum (\ref{spect}) the dependence of the critical frequencies
$\nu_{min}$ and $\nu_{max}$ on radius is determined by the magnetic
field $B(r)$. The latter can be taken from observations of $B$ in
H$_2$O maser sources by Fiebig \& G\"usten (\cite{fiebig})
(see also Liljestroem \& Gwinn 2000 for more recent measurements)

\be
\label{magmas}
B\sim 60\left({nT\over 10^{12}~{\rm cm^{-3}~K}}\right)^{1/2}~{\rm mG}.
\ee
For a standard $\alpha$-disk (see Pringle \cite{pringle}) with 
temperature

\be
\label{disktem}
T\simeq \left({3GM\dot M_d\over 8\pi\sigma r^3}\right)^{1/4},\quad
r\gg R_\ast,
\ee
and kinematic viscosity

\be
\eta\simeq \alpha c_sH\simeq \alpha c_s^2r\sqrt{{r\over GM}}, \quad
\alpha\leq 1,
\ee
one obtains

\be
\label{diskden}
\rho={\Sigma\over H}\simeq {\dot M_d\over 3\pi \eta H},
\ee
here $M$ is the mass of a central star, $\dot M_d$, the accretion rate,
$c_s$ is the sound speed in the disk, $R_\ast$ is the star radius, the other
notations have their usual meaning. When $T$ from Eq. (\ref{disktem}) and $\rho$
from Eq. (\ref{diskden}) are substituted in Eq. (\ref{magmas}), one obtains 
$b=21/16$

\be B\propto r^{-21/16}. \ee Following the above arguments we
arrive at $r\propto \nu^{-16/21}$ for $e_{\mathrm max}=$const. This is
slightly weaker than observed in the TW object, however, given the 
observational uncertainties and the rough character of these
estimates, the obtained correspondence between theoretical and
observed size--frequency dependence can be considered as
satisfactory. One should mention in this connection that a possible
deviation of $\Sigma(r)$ from that predicted for standard steady
$\alpha$-disks (see, Nelson, Benz, \& Ruzmaikina \cite{nelson}),
and radial variations of the electron spectrum may contaminate
this correspondence. It is worth noting in this regard, that
when relativistic electrons are produced by a jet impinging on a
dense ambient gas, an increase in surface brightness of the synchrotron
emission is expected in the regions where the jet produces bow
shocks and electrons are efficiently accelerated (in our case,
these are the east and west ends of the synchrotron structure).
When being accelerated in the bow shocks relativistic electrons
can diffuse and fill the jet itself; however, their energy
decreases and thus the surface brightness drops at lower radius
(see for discussion Begelman, Blandford \& Rees \cite{begelman}).
In such conditions, an inverse dependence of the angular size on
frequency, i.e. the angular size increasing with frequency, should
hold.

\section{Synchrotron structure around TW object: disk versus jet}

\subsection{Dust emission: disk}

Wyrowski et al. (1999) found that the synchrotron source in the
vicinity of the TW object is associated with a thermally radiating
dust cloud which also has a E-W elongated morphology. The
total luminosity of the cloud in the whole range of wavelengths
$\sim 4\times 10^4\lsun$ is estimated by them from an upper limit
on the contribution of the dust emission of 0.5 Jy at 220 GHz and the 
distribution of the rotation temperature of HNCO. It is obvious
that such a luminosity cannot stem from interaction of the stellar
wind, whose mechanical luminosity is $\sim 150~ \dot
M_{-6}v_8^2~\lsun$, with the disk. It is worth noting, however,
that the {\it observed} thermal dust emission in the 220 GHz range
($\sim 0.3 \lsun$ at the assumed distance to the TW object of 2.2
kpc) is only a small fraction ($\sim 2\times 10^{-3}$) of the
mechanical luminosity.

The most striking finding they mentioned is
that one of the three brightest spots of dust emission is coincident
with the center of the synchrotron source. Wyrowski et al. (1999)
argue that this coincidence
confirms that the hot dust originates from interaction of the
``synchrotron jet'' suggested by Reid et. al. (1995) with the
dense core gas confining the jet. In the disk model the observed
similarity in morphologies of the synchrotron and dust emission,
and particularly, the coincidence of their geometric centers, may
indicate that both are connected with geometry of the
dominant disk-like flow. The E-W alignment of the dust emission
and the synchrotron source, the three
brightest dust spots (A, B and C in Wyrowski et al. 1999
nomenclature), and molecular emission (such as CH$_3$CH$_2$CN,
CH$_3$OH, HCOOCH$_3$, H$_2$CO, and SO$_2$, see Fig. 3 in Wyrowski
et. al. 1999) can be understood within a picture of a circumstellar
disk embedded in a parent cloud and rotating with
angular velocity perpendicular to this direction.

\subsection{Dust emission: jet}

Reid (private communication) mentions, however, another possible
explanation: the dust giving rise to spot A (the brightest
of the three) is heated by radiation of the massive star producing the
synchrotron jet, while the other two (B and C, which are located
approximately on the continuation of the synchrotron emission to the
west) are connected with the other massive stars. Note, however, that
the hypothesis of the A, B and C spots arising due to energy radiated by 
embedded stars does not contradict the disk model itself.

\subsection{Kinematics of water masers: disk}

Fiebig et al. (\cite{fiebig96}) have shown that the maser kinematics 
admits an ambiguous interpretation: it can be explained both within 
the disk and the outflow hypothesis. We argue that this ambiguity exists 
in the case of the TW object as well.

Alcolea etal. (1993) showed that the distribution of the maser spots in 
W3(OH) consists of 3 major
clusters - one in the center plus eastern and western clusters
located at approximately the same distance from the center. This
is reminiscent of the masers which appear in the disk with the
bright source in the center and a central decrease in the
distribution of masing water. Such a situation in the close vicinity
of the TW object is quite plausible. However, careful analysis of
the situation requires thorough modelling which takes into account
special kinematics of the turbulent disc and changes in the maser
pumping conditions. This is beyond the scope of the present paper.

One possibility to tell the difference between the two hypotheses
comes from the existence of the spectral features with velocities
greatly different from the systemic velocity of the object: the maser
velocities spread from -93 km/s to -17 km/s (Cohen \cite{cohen})
while the systemic velocity is about -48 km/s. Recent sensitive
observations of the water masers in W3(H2O) displayed features for
which the difference in velocities with the bulk of the TW material
reaches 70-90 km s$^{-1}$ (Sobolev et al., private communication). 
The high velocity maser features can
appear in the clumps which are torn off the surface close to the
central parts of the disk. In this case maser spots corresponding
to blue-shifted and red-shifted features are likely to have
similar distribution in the field of view. Proper motions of the
maser features under the disk hypothesis are determined by the proper
motion of the TW object itself, the disk rotation and the turbulent
motions at the disk surface. The procedure of estimating the TW object 
proper motion should be the same under both hypotheses and it is
reasonable to assume that it is about 20 km/s as was estimated
by Alcolea etal. (1993). The sense of the disk rotation can be
guessed from the velocity gradient of the highly excited molecules
in closest vicinity of the TW object observed by Wyrowski etal. (1997): it
should be eastward for the blue-shifted and westward for
the red-shifted features. The value of the rotation velocity for the
material responsible for the high velocity features is uncertain.
However, in the central vicinity of the disk this quantity can be
considerably greater than both the proper motion of the TW object 
and the spread of the 
turbulent velocities. In this case the eastward direction of the
blue-shifted maser spots and the westward direction of the red-shifted
ones should be pronounced. The proper motion pattern in the case
of the jet is discussed in the next subsection and is greatly
different.

\subsection{Kinematics of water masers: jet}

Alcolea etal. (1993) considered water maser kinematics under the 
hypothesis of bipolar outflow and found no major contradictions.
We further consider the situation with high velocity maser
features in the case of a jet. As was mentioned in the previous
subsection, the possibility to discriminate between the two hypothesis
can be provided by the interferometric measurements of positions
and proper motions of the high velocity water maser features.

The results of Alcolea etal. (1993) show agreement with a considerable
increase of the water maser velocities toward the center. In this
case the high velocity features are expected to be situated close
to the center. Observations of Wilner et al (1999) show that the
synchrotron feature wiggles. Anyhow, the observed jet
convolution is not high and the maser kinematics shows that the
central parts of the jet are most likely moving close to
the tangential direction. In this case the proper motion of each spot
producing high velocity maser feature should be much greater than
the difference between the radial velocity of the feature and that
of the TW object. These differences for the high velocity features
considerably exceed the TW proper motion (about 20 km/s according
to Alcolea etal., 1993). So in contrast to the case of the disk
the proper motions of blue-shifted spots which trace the jet
should be always great and should have different directions:
westward for the spots to the west of the center and eastward for
the spots to the east of the center. Further, in contrast to the
case of the disk, the proper motion pattern for the red-shifted
spots which trace the jet should be the same as that for the
blue-shifted ones.

\subsection{Wiggling: disk}

Wilner et al (1999) have found that the synchrotron structure is not
precisely straight, but rather wiggles at the WE ends with a remarkable
point-like symmetry around the center.
In the disk model a wiggling synchrotron structure may reflect a 
warped structure of the disk itself, which can be produced either by 
gravitational perturbation from companion stars or sufficiently massive 
gas clumps, or by a large-scale hydrodynamical instability driven by 
radiation from the central star (Pringle 1996, Armitage \& Pringle 1997).

Wyrowski et al. (1999) estimate the star embedded in the dust spot
C to be a B0 star, and a total mass of dust emitting gas of
$\sim 15\msun$. The corresponding gravitational perturbation on
the disk from both the star and the gas in the B and C clumps can be
evaluated by comparing the vertical components of the
gravitational force from the central star $\sim MH/r^3$, where $H$
is the vertical scale height of the disk, and from a companion
star and surrounding gas $\sim (M_c+ M_{\rm BC})a/r^3$, where $a$
is the distance between the disk plane and the center of mass of
clumps B and C and an embedded star. The geometry of the TW source and of 
dust spots A, B and C suggests that $a\sim H$, which leads to the
conclusion that the perturbation from dust clumps B, C and the
star is sufficient to warp the disk if $(M_c+M_{\rm BC})\sim M$.

A complementary source for keeping the disk warped may be
connected with the interaction of the light from the central star
with the disk: when the intrinsic luminosity of a central star
settling onto the main sequence overwhelms the total disk
luminosity, it can initiate a large scale instability and bend the
disk (Pringle 1996, Armitage \& Pringle 1997). The threshold
luminosity for the instability to grow is $\sim 10L_\odot$, and is
much less than the total luminosity of the stars embedded in the TW
structure, $L_{\rm TW}\sim 3\times 10^4L_\odot$ (Wyrowski et al.
1999). The warp excited initially at the inner disk diffuses outside
and can last tens of Myrs. At later times, $t=15$ Myr, the tilt
angle between the inner ($r\sim 10$ AU) and outer ($r\sim 100$ AU)
disk can reach $\sim 10^\circ$ (Armitage \& Pringle 1997).
Armitage \& Pringle argue that this mechanism is most likely
responsible for warping of the inner disk in $\beta$ Pictoris
(Burrows et al. 1995). As shown by Quillen (2001) a strong stellar
wind is also expected to produce disk warping.

A change of the wiggle morphology in the disk model can appear
either due to relative motions of the central star (traced
presumably by the A spot) and its companions (traced by B and C
spots) if warping is produced by gravitational perturbation from
these sources, or by proper rotation of the disk. In both cases
the characteristic times are much longer than in the jet model:
$t\sim \sqrt{GM/r^3}\sim 5\times 10^3$ yr for $M\sim 10\msun$, and
$r\sim 0.01$ pc. This is consistent with observed lack of
morphology variations over 510 days (Wilner et al. 1999).

\subsection{Wiggling: jet}

Wilner et al (1999) argue that wiggling of the synchrotron source
in the TW may be the result of precession of the jet axis. In principle,
precession is typical for jets in the vicinity of binary
stellar objects and in AGNs. In the TW object precession can be
maintained also by the gravitational influence from a companion
star associated with the B and C dust spots. Large scale wiggling
can be connected with a long-periodic precession with
characteristic time $t\sim L/u$, where $L$ is the length of the jet,
and $u$ is the speed of flow within it. The synchrotron structure in
the TW object has an extent $2L\sim 0.01$ pc (Reid et al. 1995), and
the velocities of the water masers provide a lower estimate of the
flow velocity: $u\sim 20$ km s$^{-1}$ at 1$''$ and increasing to
the center (Alcolea et al. 1993), $u\simlt 40$ km s$^{-1}$ (Cohen
1979). Recent sensitive observations of the water masers in
W3(H2O) displayed features for which the difference in velocities
with the bulk of the TW material reaches 70-90 km s$^{-1}$ (Sobolev
et al., private communication). Thus, the
upper limit for long-periodic variations of the wiggle morphology
is 250 yr, while the lower limit can be not less than five years
for the jet flow slower than 1000 km s$^{-1}$ expected for the
conditions in the vicinity of a young OB star. These estimates do
not contradict the data of Wilner et al. (1999) on the absence of
noticeable variability within 510 days.

\subsection{Polarization: disk}

The jet and the accretion disk surface are expected to show different
polarization properties. In the first case the emission is
expected to be linearly polarized because it arises in the regions with
an apparently stronger and rather regular magnetic field of the jet.
In principle, the interface between the jet and the ambient medium is subject
to the Kelvin-Helmholtz instability, which forces the interface to settle
into a regime with highly developed turbulent motions causing
depolarization. Indeed, numerical simulations show that super-Alfv\'enic
jets undergo KH instabilities; however, only the relatively narrow region
close to the jet walls becomes involved in a strongly nonlinear turbulent
regime, while the body of the jet remains more regular since 
the characteristic growth
rate of the instability there is an order of magnitude smaller than at the
interface. As a result, the modelled synchrotron emission reveals a high
degree of polarization (Hardee \& Rosen 1999).

Wilner et al. (1999) mention that morphologically the synchrotron
structure in the TW source is similar to that observed in AGNs, where
polarization in the jet component can reach 10\%. In some cases
the degree of polarization in extragalactic jets can be as high as
50 \% -- an upper limit for optically thin synchrotron emission
(Perlman et al., 1999), which is consistent with the predictions of
Hardee \& Rosen (1999).

In the case of synchrotron radiation produced in the boundary
layer of the disk, where the acceleration of non-thermal electrons is
intimately connected with a highly developed turbulence, the
emission seems to be depolarized. Wilner et  al (1999) reported
that the synchrotron source in the TW does not show polarization
higher than 0.05 of the total intensity of the core of the
synchrotron structure, and from this point of view it is
consistent with the disk model.

\subsection{Polarization: jet}

However, Wilner et al. bring forward several arguments explaining 
the nondetection
of polarization from the TW synchrotron source. The fact that the core
component does not show significant polarization might reflect the
similarity between this source and the jets in ANGs where the core
component emission shows typically less than a few percent of the
total emission. Note, that in the case of the TW source it can be connected
with a short periodic (a few days) variations of the core
component produced by a strong influence of the magnetic
field of the magnetosphere of the central star on the foot region
of the jet, and highly developed random motions caused by this 
(see, e.g. Lovelace, Romanova, \& Bisnovatyi-Kogan 1995, Goodson,
B\"ohm, \& Winglee 1999). The jet component, where AGN jets normally
show polarization, in the case of the TW source is too weak for the confident
detection of polarization.

On the other hand, internal Faraday depolarization produced by thermal
electrons with $n_e\simgt 10^3$ cm$^{-3}$ would be sufficient to 
explain the low level of polarization. The upper limit derived by 
Wilner et al. for $n_e$ from the lack of detected thermal emission 
in the cm-wavelength range and from the size of the TW synchrotron structure 
is more than two orders of magnitude larger than this value. 
Under these circumstances a more precise determination of the thermal electron
density would be of critical importance.


\section{Conclusions}

   \begin{enumerate}
      \item We have shown that synchrotron emission observed from the TW
object can, in principle, be connected with the accretion disk
around a massive star. Interaction of the wind and the strong
shocks developed in the wind flow with the disk surface can
generate multiple shocks in the boundary layer on the disk
surface, which in turn can efficiently produce relativistic
electrons. If only as little as than 0.5\% of the wind kinetic energy is
transformed into relativistic electrons, the total synchrotron
luminosity can be comparable to that observed around the TW object 
in the W3(OH) region.
      \item The observed dependence of the 
      angular size of the source on frequency
can be explained as due to radial variation of the upper cutoff frequency
of the synchrotron spectrum, which can be connected in particular with
the radial dependence of the magnetic field in the disk.
      \item We have shown that existing data on the spatial
distribution of dust emission, kinematics of molecular material as
well as morphology and polarization of the synchrotron emission
around the TW object can be explained both with the disk and the 
jet hypothesis. Further observational efforts in the directions shown
(e.g., measurement of proper motions of the water maser
spots corresponding to the high velocity features) can
distinguish between the two hypotheses.
   \end{enumerate}

In this paper we did considered synchrotron emission only around the TW
object. Recently there have been reports of other sources of combined 
radiocontinuum and water maser emission associated with massive young 
stellar objects which probably contain disks (see Menten \& van der
Tak, 2003 and references therein). The bulk of the 
radiocontinuum emission around these sources looks different from
synchrotron emission as can be inferred from their spectral indices. 
Thus, the suggested phenomenon can be widespread
in the massive objects but difficult to observe because of
confusion with other types of emission.

In our model the synchrotron luminosity of the circumstellar disk is
proportional to the mass loss rate $\dot M$, the wind velocity $v$, and the
disk radius $R$; it depends on the magnetic field slightly more strongly, as 
$B^{1.6}$. 
One can thus think that only the disks around massive stars with high
enough $\dot M$ and $v$ can be sufficiently bright in synchrotron 
radiation. Note,
however, that among the essential factors which can determine
synchrotron luminosity are also the separation between the shock waves $l_s$,
and the spectrum of the turbulent magnetic field $\sim k^{-\xi}$; additional
study is needed to draw firm conclusions on whether disks around less
massive stars, such as for instance T Tauri, can shine in synchrotron 
radiation.

\begin{acknowledgements}
      We thank the referee M. J. Reid and K. Menten for valuable critical comments.
      Stimulating discussions with I. I. Zinchenko are greatly
      acknowledged.
      Part of this work was supported by INTAS project 99-1667, RFBR projects 02-02-17642 and 03-02-16433, and grant of Ministry of Education of Russian Federation. YS acknowledges financial
      support from \emph{Deut\-sche For\-schungs\-ge\-mein\-schaft, DFG\/}
      (project SFB N 591, TP A6).
\end{acknowledgements}

\end{document}